Pharm Res (2019) 36:116
https://doi.org/10.1007/s11095-019-2643-2RESEARCH PAPER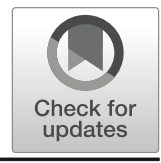

# Natural Deep Eutectic Solvents as Agents for Improving Solubility, Stability and Delivery of Curcumin

Tomasz Jeliński [1] • Maciej Przybyłek [1] • Piotr Cysewski [1]Received: 8 November 2018 / Accepted: 6 May 2019
© The Author(s) 2019## ABSTRACT

**Purpose** Study on curcumin dissolved in natural deep eutectic solvents (NADES) was aimed at exploiting their beneficial properties as drug carriers.
**Methods** The concentration of dissolved curcumin in NADES was measured. Simulated gastrointestinal fluids were used to determine the concentration of curcumin and quantum chemistry computations were performed for clarifying the origin of curcumin solubility enhancement in NADES.
**Results** NADES comprising choline chloride and glycerol had the highest potential for curcumin dissolution. This system was also successfully applied as an extraction medium for obtaining curcuminoids from natural sources, as well as an effective stabilizer preventing curcumin degradation from sunlight. The solubility of curcumin in simulated gastrointestinal fluids revealed that the significant increase of bioavailability takes place in the small intestinal fluid.
**Conclusions** Suspension of curcumin in NADES offers beneficial properties of this new liquid drug formulation starting from excreting from natural sources, through safe storage and ending on the final administration route. Therefore, there is a possibility of using a one-step process with this medium. The performed quantum chemistry computations clearly indicated the origin of the enhanced solubility of curcumin in NADES in the presence of intestinal fluids. Direct intermolecular contacts leading to hetero-molecular pairs with choline chloride and glycerol are responsible for elevating the bulk concentration of curcumin. Choline chloride plays a dominant role in the system and the complexes formed with curcumin are the most stable among all possible homo- and hetero-molecular pairs that can be found in NADES-curcumin systems.

**KEY WORDS** curcumin · delivery · FASSIF · NADES · solubility## ABBREVIATIONS

FASSGF    Fasted State Simulated Gastric Fluid
FASSIF    Fasted State Simulated Intestinal Fluid
NADES    Natural Deep Eutectic Solvent

## INTRODUCTION

Curcumin is a polyphenol occurring naturally in the rhizome of the herb *Curcuma longa*, known popularly as turmeric in India. From a chemical point of view it is a α,β-unsaturated diketone (diferuloylmethane) exhibiting keto-enol tautomerism. Since ancient times it has been used along other curcuminoids extracted from turmeric in traditional medicine due to its beneficial influence on human health. Curcumin shows anti-oxidant (1), anti-inflammatory (2), anti-microbial (3) and anti-tumor (4), nephro-protective (5) and anti-diabetic (6) activities. Importantly, curcumin is safe even at high doses (7). Despite these properties, clinical application of curcumin is limited due to its low water solubility and bioavailability, as well as susceptibility to degradation, particularly in the presence of light (8). Multiple strategies have been employed to overcome these issues (9–13), however none of them have been completely successful.

The delivery of a drug in the human body is aimed at achieving or optimizing the therapeutic effects of the drug, while minimizing its adverse effects (14). Applicability and bioavailability of active pharmaceutical ingredients (APIs) requires their dissolution in aqueous media at some point of the administration rout of the drug (15). Therefore, the problem of limited aqueous solubility of drugs becomes an important issue, since as studies show a considerable part of newly developed drugs can be regarded as practically insoluble (16). This poses a need for developing new formulation strategies overcoming the problem of poor solubility of APIs. The importance of this challenge lies in the fact that limited solubility does not only hinder the bioavailability of a drug, but can also result in significant side effects. These side effects are a consequence of dose escalation, which is necessary in the case of poorly soluble drugs, and manifest particularly in the irritation of the gastrointestinal tract (17). Several different delivery strategies were developed to address the problem of inadequate solubility of APIs and they can be divided into two general types of strategies. The first group is focused on the reduction of particle size of the drug and includes micronization (18) or forming nanocrystals (19). The second group utilizes the change of the crystal form of the drug, which can include cocrystals formation (20), formation of salts (21) or creating metastable polymorphs (22). There are also other strategies aimed at increasing drug solubility, namely complexation with cyclodextrin (23), amorphization (24), formulation of solid dispersions (25), modification of pH (26) or solubilization in lipids (27). However, it would be also advantageous to achieve a desired therapeutic effect with the use of a simple solvent without the costly and time consuming procedures. This requires a search for new media, which would offer both higher solubility of the active ingredient, as well as environmental and health safety. In this context, a particularly interesting group of solvents are natural deep eutectic solvents (NADES).

✉ Tomasz Jeliński
tomasz.jelinski@cm.umk.pl

[1] Chair and Department of Physical Chemistry, Faculty
of Pharmacy, Collegium Medicum of Bydgoszcz, Nicolaus Copernicus
University in Toruń, Kurpińskiego 5, 85-950 Bydgoszcz, PolandPublished online: 03 June 2019🙆 Springer

In general, deep eutectic solvents (DES) can be treated as mixtures of solids compounds that can form liquid solutions even at room temperature, which is a result of lowering of mixture melting point (28). In particular, natural deep eutectic solvents (NADES) are a specific subgroup of DES comprising such primary plant-based metabolites as organic acids, alcohols, amino acids or sugars (29). There are several beneficial properties of NADES which distinguish them from traditional solvents and these include low volatility, liquid state even in sub-zero temperatures, biodegradability, sustainability, high extraction and stabilization potential, low cost and simplicity of preparation (30). It is then no surprise that natural deep eutectic solvents have found their application in many different areas, such as extraction (31), stabilization (30), biotransformation (32), enzyme reactions (33), improving solubility in the aqueous phase (34) or enhancement of bioactivity (35).

This particular study was focused on the search for an optimal NADES formulation that would allow for an enhanced solubility and bioavailability of curcumin. For this purpose several NADES constituents that are already well-established in the literature were used, namely sugars and sugar alcohols (36,37) Especially interesting was the behavior of curcumin in simulated gastrointestinal fluids, acting as models of natural conditions of drug absorption. The general premise of the study was that natural deep eutectic solvents could be employed throughout the whole process of curcumin handling, *i.e.* from the extraction from natural sources up to oral administrations to individuals. Also, the COSMO-RS computational protocol, which was already used in the context of NADES properties (38), was utilized for a deeper understanding of the behavior of the analyzed systems.

## MATERIALS AND METHODS

### Materials

Curcumin (CAS: 458–37-7) was provided by Sigma Aldrich. Natural deep eutectic solvent constituents comprised eight different compounds, namely choline chloride (CAS: 67–48-1), glucose (CAS: 50–99-7), fructose (CAS: 57–48-7), sorbitol (CAS: 50–70-4), xylitol (CAS: 87–99-0), maltose (CAS: 69–79-5), sucrose (CAS: 57–50-1) and glycerol (CAS: 56–81-5), which were also purchased from Sigma-Aldrich. Methanol, used as a solvent, was obtained from Avantor Performance Materials, Poland. All chemicals were purchased at their highest available purity and were used as received, apart from choline chloride which was dried before use. Fasted State Simulated Intestinal Fluid (FASSIF) and Fasted State Simulated Gastric Fluid (FASSGF) powders were purchased from Biorelevant.com Ltd. and were prepared according to the instructions. Commercial turmeric powder was purchased from three different manufacturers and used for curcumin extraction without any modifications.

### Methods

#### NADES Solubility Measurement

Natural deep eutectic solvents were prepared by mixing choline chloride together with one of the other NADES components. Weighed amounts of these mixtures were sealed in test tubes and heated at 80°C in a water bath until homogenous solutions were formed. In order to determine the solubility of curcumin in a specific NADES its excess amount was added to the test tube containing the NADES composition which was then stirred and allowed to reach equilibrium conditions at room temperature. After 24 h the compositions were centrifuged ensuring that the remaining excess amount of curcumin was left on the bottom of the test tube. Before the actual solubility measurements the solution was filtered using PTFE syringe filter with a 0.20μm membrane. A small amount of NADES supernatant containing the drug was then weighed and diluted in methanol. Samples were diluted in order to obtain a value of absorbance within the linearity limit of the calibration curve. The concentration of the drug in NADES was measured spectrophotometrically on a Biosens UV-6000 spectrophotometer and the absorbance was measured at 422 nm. The concentration of curcumin was based on a calibration curve (see Fig. 1) obtained by dissolving pre-weighed amounts of curcumin in methanol and measuring their absorbance in the function of concentration, expressed as mass of the drug dissolved in an amount of the solvent. Three samples were measured for each of the considered systems. The influence of temperature on the solubility of curcumin in NADES was tested by heating the formulation and adding an excess amount of the drug to a test tube. The system containing curcumin in NADES was then kept in a water bath at controlled temperature for 1 hour. The final procedure of determining the concentration was identical as the one described earlier.

### Extraction Measurements

Turmeric powder from three different commercial suppliers was extracted with water, methanol and NADES comprising choline chloride and glycerol. An excess amount of the powder, that is 250 mg, was placed in a sealed vessel together with an appropriate solvent and then placed on a mechanical stirrer. After 24 h the concentration of curcumin in the solution was measured according to the procedure described earlier.

### Stability Measurements

Two samples of curcumin, one dissolved in methanol and the other in NADES, were prepared and their concentrations were measured. These samples were exposed to artificial sunlight and after fixed intervals of time the concentration of curcumin in the samples was determined. Also powdered curcumin was tested and in this case the procedure involved the dissolution of a specific amount of curcumin powder in methanol and then the determination of its concentration. An artificial sunlight lamp was used for simulating the effect of sunlight exposure. The power density of this lamp was measured as a function of the length from the lamp to the surface containing the examined epoxy pellets. Also, the temperature was measured in order to avoid overheating of the samples.

The length of 12.5 cm from the samples was chosen, giving a power density of 9.2 kW/m$^2$. A series of time intervals from 15 to 120 min was used for exposure of the samples and the results were presented in the function of the calculated irradiance, expressed in kWh/m$^2$.

### FASSIF and FASSGF Solubility Measurements

Fasted State Simulated Intestinal Fluid (FASSIF) and Fasted State Simulated Gastric Fluid (FASSGF) were used to determine the solubility of curcumin in gastrointestinal fluids. These two simulated fluids were prepared according to the instructions provided by the producer and the curcumin containing NADES formulation was prepared according to the procedure described earlier. Different volumes of simulated gastrointestinal fluids were used to reflect their different amounts among individuals. A fixed amount of a NADES formulation weighing 300 mg and containing 2.3 mg of curcumin was added to the fluids. After the precipitation of the excess amount of curcumin, a sample was taken from the fluid and centrifuged ensuring that the remaining excess amount of curcumin was left on the bottom of the test tube. A small amount of the fluid containing dissolved curcumin was then measured spectrophotometrically and after the measurement the whole sample, including the undissolved curcumin, was returned to the main volume of the fluid. A series of gradual dilutions was prepared by adding different amounts of simulated fluids and for each dilution the concentration of curcumin was measured. For comparison, the solubility of curcumin powder in FASSGF and FASSIF was measured, similarly as in the case of curcumin solubility in NADES.

### Quantum Chemistry Computations

Structures of choline chloride, glycerol and curcumin were represented by series of the most probable conformations generated using COSMOconf 4.2 (39) on COSMO-BP/def2-TZVPD-fine level. Thermodynamic characteristics were described with an aid of COSMOtherm Version 18 (40),

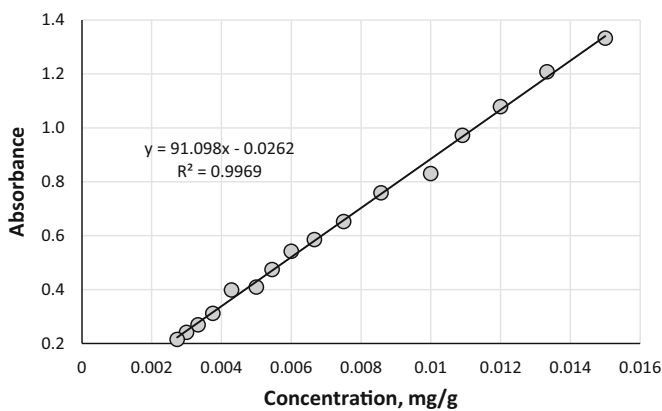

**Fig. 1** Calibration curve for curcumin, obtained by plotting the absorbance of the drug in the function of its concentration in methanol solvent.

$y = 91.098x - 0.0262$
$R^2 = 0.9969$

using the BP-TZVPD-FINE_18.01 parameterization. Interactions of components in NADES were described by inclusion of homo- and hetero-molecular pairs. Bimolecular clusters were obtained according to the contact probability computed using COSMOtherm facilities. Obtained geometries were pre-optimized with an aid of COSMO-BP/SVP approach and all resulting structures of relative energies higher than 25 kcal/mol were rejected. The remaining pairs were re-optimized using COSMO-BP/def2-TZVPD-fine approach. Thermodynamic characteristics of curcumin-NADES solutions was based on the Gibbs free energy of a $X + Y = XY$ reaction type, where X,Y denote the considered species. Since the energy of the compounds in the gas phase has the strongest influence on the accuracy of the equilibrium constants and components affinity, the contributions coming from electron correlation were included, using RI-MP2 *ab inito* approach. The def2-QZVPP basis sets were used as implemented in Turbomole 7.0 (41). Also, for a proper accounting of the rotational and translational enthalpy in the gas phase to the enthalpy part of thermal energy, the vibrational zero-point energy (ZPE) of all the species was computed. It was advised (42) that both ZPE + RI-MP2/def2-TZVPP methods provide reasonable accuracy of Gibbs free energy calculation.

## RESULTS AND DISCUSSION

### Solubility of Curcumin in NADES

The solubility of curcumin in seven different natural deep eutectic solvents was measured spectrophotometrically based on the calibration curve documented in Fig. 1, which provided an acceptable linear correlation for calculating the concentration of curcumin in the solution, with the coefficient of determination equal to almost 0.997.

The tested NADES varied not only in their chemical composition, utilizing choline chloride together with seven different constituents as listed in Table I, but also in their molar proportions. Three different ratios were used, namely a unimolar proportion and a threefold excess amount of both choline chloride and the second component, which yielded a total of twenty one different combinations. The obtained solubilities, expressed as the mass concentration of the drug in NADES, were compared to the solubility of curcumin in water, which was determined by experiment to be equal to 0.0006 mg/g. This led to the calculation of the solubility advantage of the studied systems, defined as the ratio of the solubility of curcumin in NADES and in the aqueous solution. The obtained solubility values, together with the calculated solubility advantage, were collected in Table I. It is evident that all of the studied formulations were responsible for a significant increase of the solubility of curcumin compared to water. Among these systems the NADES comprising choline chloride and glycerol gave the largest increase in curcumin solubility, regardless of the molar ratio used. In particular, for the unimolar proportion, which was found to be the most successful in terms of solubility increase, the solubility of curcumin was found to be 7.25 mg/g, which amounts to a solubility advantage of more than twelve thousand.

As for other NADES, the systems with fructose and glucose gave the second most significant solubility increase, *i.e.* around 5000 compared to water. Using sorbitol and xylitol as the second NADES component resulted in an even smaller solubility advantage, although its value was still as high as 2000. The smallest solubility increase, around 1000, was observed for systems utilizing sucrose and maltose. Certain regularities in the influence of the molar ratio of constituents were observed. For the first five most efficient NADES the highest solubilities were obtained for the unimolar proportion. Furthermore, for the three most successful ones the second best composition involved an excess amount of the second component. In the case of the other two systems, *i.e.* utilizing sorbitol and xylitol, the second optimal proportion involved an excess amount of choline chloride. As for the two least successful NADES, this order was different since the solubility of curcumin increased with the increased amount of choline chloride in the composition.

**Table I** Solubility At 20°c of Curcumin (Mg/G) in Natural Deep Eutectic Solvents Comprising Choline Chloride Combined With Selected Sugars and Sugar Alcohols. Values of Solubility Advantage Are Provided In Brackets

| NADES constituent | Molar ratio of choline chloride and the second NADES constituent | | |
|---|---|---|---|
| | 3:1 | 1:1 | 1:3 |
| Glycerol | 2.32 (3870.0) | 7.25 (12,089.7) | 3.15 (5255.2) |
| Fructose | 1.02 (1706.3) | 3.11 (5179.9) | 0.79 (1319.9) |
| Glucose | 0.99 (1647.0) | 2.90 (4837.1) | 0.74 (1226.1) |
| Sorbitol | 0.38 (641.3) | 1.29 (2148.2) | 0.44 (725.3) |
| Xylitol | 0.33 (558.2) | 1.17 (1954.6) | 0.37 (628.6) |
| Sucrose | 0.65 (1089.3) | 0.27 (444.0) | 0.20 (340.2) |
| Maltose | 0.61 (1010.3) | 0.21 (345.7) | 0.17 (285.6) |

The most promising NADES system, that is the one utilizing choline chloride and glycerol, was selected for further tests involving an extended range of molar compositions and different temperatures. The range of molar ratios of constituents was extended by an addition of five- and seven fold excess amounts of either of the components with the dissolution temperature ranging from 20°C to 60°C with a 10 degree interval. The obtained results were collected in Table II.

The increased temperature resulted in raised curcumin solubility values for all of the studied molar compositions. For the unimolar proportion, responsible for the largest solubility of curcumin, the temperature increase from 20 °C to 60°C gave a solubility raise of less than 20%. Interestingly, this was the smallest increase in solubility among all of the studied systems. It was found, that increasing either the amount of choline chloride or glycerol results in a more pronounced solubility gain. For a sevenfold excess amount of choline chloride this solubility gain was the largest and amounted to around 120%.

**Extraction of Curcuminoids from Turmeric Powder**

The system which turned out to be the most effective in terms of curcumin solubility was selected for studies on the extraction of curcuminoids from turmeric powder. NADES comprising choline chloride and glycerol in unimolar proportions was tested along with methanol and water for reference. Three commercial turmeric powders from different suppliers were used and the results of the tests were collected in Table III.

The three different turmeric powders varied only slightly and gave very similar results in the case of each solvent. Methanol turned out to be the most effective extraction solvent among studied media, as the extracted amount of curcuminoids reached more than 5% of the initial mass of the turmeric powder sample. On the other hand, this value was minimal in the case of water. The chosen NADES was responsible for the extraction of around 1% of the turmeric powder mass. Although the extracted amount was smaller than for methanol, it was still a significant one and the advantage of using natural deep eutectic solvents for this role lies in the fact that they are safe to use in terms of health hazards caused by organic solvents traditionally used for extraction of curcuminoids. Besides, it is known that that there is synergistic effect of three major curcuminoids present in natural herb *Curcuma longa*, what was even patented and marked under trade mark Curcumin C3 complex®. Therefore, the application of NADES as an extraction media also takes advantage of this natural mixture of three major curcuminoids.

**Stability of Curcumin in NADES**

The stability of curcumin is important in the context of preserving its pharmaceutical activity but also because curcumin can be used as a dye and therefore requires adequate color stability. It is known that curcumin is sensitive to light and temperature which limits its applications (43). This was the motivation for the tested involving exposure of curcumin sample to artificial sunlight which would simultaneously simulate the effect of light and temperature. Curcumin dissolved in methanol and NADES was used, together with curcumin powder. The details of the procedure can be found in the methodology part and the results were presented in Fig. 2.

Curcumin in the form of powder and dissolved in methanol experienced a significant degradation after exposure to artificial sunlight. In the case of methanol solution the concentration of curcumin in the sample dropped to 5% of its initial amount after 120 min. For curcumin powder this decrease was less pronounced although after the same time only 19% of curcumin remained in the sample. In contrast, curcumin dissolved in NADES, choline chloride and glycerol in unimolar proportions, remained stable through the whole time of exposure and no degradation was observed. This is a particularly interesting observation in regard to the significant extraction potential of the tested NADES which makes it possible to safely store curcumin in this solvent immediately after extraction from natural sources.

**Table II** Solubility of Curcumin (Mg/G) In Natural Deep Eutectic Solvents Comprising Choline Chloride And Glycerol In Varying Temperatures. Values of Solubility Advantage Are Provided In Brackets

| T [°C] | Molar ratio of choline chloride and glycerol | | | | | | |
|---|---|---|---|---|---|---|---|
| | 7:1 | 5:1 | 3:1 | 1:1 | 1:3 | 1:5 | 1:7 |
| 20 | 0.63 (1047.4) | 1.09 (1814.0) | 2.32 (3870.0) | 7.25 (12,089.7) | 3.15 (5255.2) | 2.56 (4274.6) | 1.94 (3238.0) |
| 30 | 0.75 (1250.0) | 1.46 (2433.3) | 2.51 (4183.3) | 7.44 (12400) | 3.32 (5533.3) | 2.72 (4533.3) | 2.02 (3366.7) |
| 40 | 1.03 (1719.5) | 1.66 (2770.5) | 2.97 (4954.7) | 7.76 (12,936.3) | 3.54 (5893.0) | 2.91 (4843.6) | 2.34 (3896.1) |
| 50 | 1.2 (2007.6) | 1.92 (3205.4) | 3.15 (5243.7) | 8.15 (13,581.8) | 4.01 (6679.9) | 3.06 (5103.5) | 2.46 (4098.1) |
| 60 | 1.38 (2303.4) | 2.11 (3513.4) | 3.65 (6080.8) | 8.6 (14,333.1) | 4.15 (6921.3) | 3.14 (5228.4) | 2.63 (4382.5) |

**Table III** Values of Concentration of Curcuminoids (Mg/G) and the Extracted Mass Percentage of Turmeric Powder. The "NADES" Designation Refers To the Most Optimal of the Studied Formulations, Designations From 1 To 3 Refer To Different Commercial Suppliers

| Turmeric powder | Methanol | Water | NADES |
| --- | --- | --- | --- |
| "1" | 2.55 (5.09%) | 0.036 (0.07%) | 0.56 (1.10%) |
| "2" | 2.64 (5.21%) | 0.041 (0.08%) | 0.64 (1.26%) |
| "3" | 2.55 (5.05%) | 0.031 (0.06%) | 0.47 (0.94%) |

## Solubility of Curcumin in FASSGF and FASSIF

The bioavailability of curcumin, and any other drug, is limited by its solubility in gastrointestinal fluids. Therefore, a series of tests was conducted aimed at determining the solubility of curcumin in Fasted State Simulated Intestinal Fluid (FASSIF) and Fasted State Simulated Gastric Fluid (FASSGF). Curcumin was used both in the form of a powder, as well as contained in NADES comprising choline chloride and glycerol in unimolar proportions. It is known that gastrointestinal fluids show significant condition dependent volumes, varying also among individuals. In particular, according to the study by Schiller *et al.* (44) the volume of the fasted gastric fluid varies from 13 mL to 75 mL (median: 47 mL), while for the small intestinal fluid from 45 mL to 319 mL (median: 83 mL). Hence, taking this into account requires different amounts of simulated fluids for mimicking the range of possible physiological conditions. Thus, the volumes of both FASSGF and FASSIF used in further solubility studies ranged from 10 mL up to 300 mL. The solubility of curcumin powder in the two simulated gastrointestinal fluids was also determined and used as a reference. Performed measurements for FASSGF the solubility of curcumin resulted in 0.001 mg/g, while for FASSIF its solubility was found to as rich as 0.004 mg/g. Analogically, the solubility of curcumin contained in NADES was determined in different volumes of gastrointestinal fluids, in the range occurring among different individuals, and the results were presented in Fig. 3. For comparative purposes also the solubility in water was provided.

After the addition of curcumin contained in NADES to the gastrointestinal fluid an undissolved excess amount of curcumin appeared in the solution and the concentration of curcumin decreased with the increased amount of the fluid. However, one could expect that if the NADES played no role in curcumin solubility the concentration would remain on the same level (0.001 mg/g for FASSGF and 0.004 mg/g for FASSIF) as the undissolved excess curcumin would be gradually dissolved by increased amount of the gastrointestinal fluid. However, this was not the case and what should be attributed to the fact that for small volumes of the internal fluids the added NADES remains a significant portion of the entire volume. For example, in 10 mL of the gastrointestinal fluid the added NADES composition, equal to 300 mg, is around 3% of the entire volume, while in 300 mL it is reduced to around 0.1%. Therefore, in small volumes of intestinal and gastric fluids the NADES remains active as an agent increasing the solubility of curcumin. This univocally points out that fasted states, corresponding to small volumes of gastrointestinal fluids, are preferable in this context. It is also evident that the solubility of curcumin in the Fasted State Simulated Intestinal Fluid is significantly larger than in the case of Fasted State Simulated Gastric Fluid, not to mention in water. For the median volumes of gastrointestinal fluids mentioned above, the concentrations were determined as follows: in 47 mL of FASSGF – 0.0035 mg/g; in 83 mL of FASSIF – 0.008 mg/g. This observation on the other hand, suggests that the small intestinal fluid is the desired place of curcumin delivery. Also, a significant gain in solubility is observed when comparing these values to the concentration of powdered curcumin in gastrointestinal fluids, *i.e.* for FASSGF the solubility of curcumin contained in NADES is around 3.5 times larger and for FASSIF it is around 2 times larger. These values demonstrate the feasibility of using natural deep eutectic solvents as curcumin solubility enhancing agents for administration to individuals.

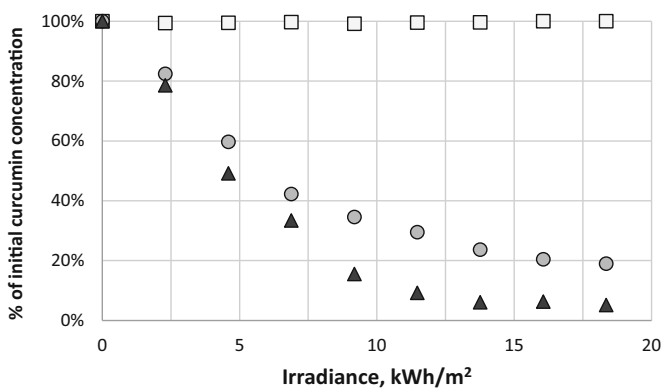

**Fig. 2** Percentage of the initial concentration of curcumin as the function of the irradiance from artificial sunlight. White squares denote curcumin dissolved in NADES, black triangles denote curcumin dissolved in methanol, while grey circles stand for curcumin powder.

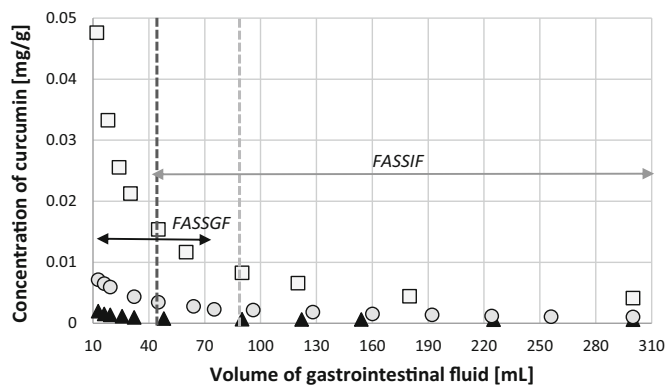

**Fig. 3** Concentration of curcumin contained in NADES in different solvents: water (black triangles), FASSGF (grey circles) and FASSIF (white squares) as the function of the volume of the simulated gastrointestinal fluid. The black vertical line denotes the median volume of the gastric fluid (47 mL), while the grey vertical line is the median volume of the small intestinal fluid (83 mL), both in fasted state. The horizontal lines depict the volume ranges of the two gastrointestinal fluids found among different individuals.

## Origins of the Enhanced of Curcumin Solubility in NADES

The observations presented above clearly suggest that all studied NADES are very effective solubility enhancers of curcumin. In order to find the origin of this fact, the detailed mechanism was studied in the case of the most effective NADES, comprising choline chloride and glycerol. Hence, advanced quantum chemistry computations were undertaken for inspection of the structure of systems comprising curcumin (A), choline chloride (C) and glycerol (B) with experimental proportions.

As mentioned in the methodology part, the affinity of components was expressed by values of Gibbs free energy of pairs formation, $X + Y = XY$, where $X,Y = \{A,B,C\}$. Hence, three dimers representing homo-molecular pairs were considered what was augmented by three hetero-molecular contacts. Schematic representation of geometries and electrostatic densities of these pairs are presented in Fig. 4.

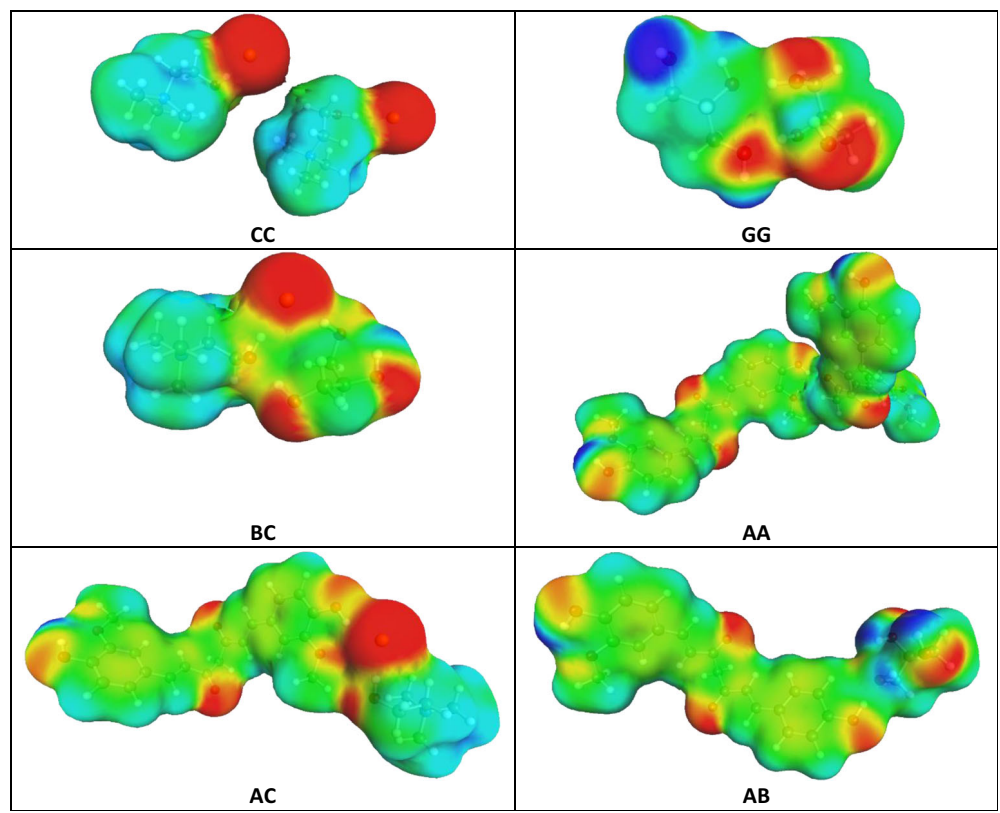

**Fig. 4** Electrostatic density contour plots of the most stable pairs of NADES-curcumin systems. Red color represents electro-negative centers, while blue color stands for electro-positive regions.

It is worth mentioning that for actual computations several conformers of either monomers or bi-component clusters were used. For illustration purpose only the most stable pairs were included in Fig. 4, which exemplifies the structural diversity of NADES-curcumin systems. All components are prone to hydrogen bonding formation or strong electrostatic interactions. In the case of curcumin the most favourable dimer is stabilized by hydrogen bonding of vicinal hydroxyl group with one of the central keto-groups. Contributions of all other structures to the overall stabilisation of the system can be inferred from Fig. 5, documenting the computed values of Gibbs free energies of reactions mentioned earlier. In the left panel of Fig. 5 the temperature influence on the properties of the 1:1 NADES-curcumin system is described. It is evident that the temperature condition has almost a monotonous influence on all values of $\Delta G_r$. Since all reactions leading to homo- and hetero-molecular complexes are endothermic and entropy driven, the rise of temperature is associated with the increase of affinity between constituents of all considered systems. This is in good accord with the observed rise of solubility of curcumin imposed by temperature elevation. Besides, temperature has only a marginal influence on the relative stabilization of complexes and the strongest stabilization is observed for hetero-molecular AC complex. Hence, in all temperatures the strongest binding of choline chloride with curcumin is responsible for the solubility alteration of this API in the studied NADES. Choline chloride is also able to form very strong dimers, which are only slightly less stable then AC complexes. The affinities of glycerol for dimerization and binding with curcumin are very comparable and lower than the two former complexes by about 2 kcal/mol. The lowest value of $\Delta G_r$ has been found in the case of self-association of curcumin and glycerol interactions with choline chloride. Hence, the relative stabilization observed in 1:1 NADES-API system can be summarized as follows: AC ≈ CC > AB ≈ BB > AA ≈ BC in all temperatures, including normal body conditions.

The right panel of Fig. 5 presents the trends of $\Delta G_r$ computed at room temperature as a function of choline chloride mole fraction in NADES. It is worth emphasizing that changes in choline chloride concentration significantly alter the properties of the obtained ionic liquid due to polarity alterations and the most significant influence can be observed in the case of the most dominant intermolecular interactions. First of all, what was expected, the increase of choline chloride concentration is responsible for an increase of stabilization of both AC and CC complexes. Furthermore, variation of choline chloride amount in the system has a very strong influence on self-association of curcumin and the rise of choline chloride share significantly reduces the stability of curcumin dimers in the solution. Since this trend is very sharp, the overall concentration of bound curcumin decreases with the increase of choline chloride in the system. This is even strengthened by a systematic decrease in stabilization of AB pairs. The mean value of intermolecular interactions is affected by choline chloride concentration as plotted in Fig. 5. It turns out that the minimum value equal to −9.1 kcal/mol has been found for the 1:1 NADES-curcumin system. This is in good correspondence with the experimentally observed solubility of NADES of varying proportions of choline chloride and glycerol. Besides, it is observed that the rise of temperature does not shift the location of this minimum and only to −9.7 kcal/mol at 60°C, which was the highest studied temperature.

In conclusion, it is worth summarizing that the computed values of $\Delta G$ point out on the source of solubility enhancement of NADES, which is directly related to interactions of curcumin with either choline chloride or glycerol. The former predominates and has the dominant contribution to the presence of bound form of curcumin in the solution. Curcumin dimers are supposed not to contribute to the solubility increase due to much lower stability compared to hetero-molecular pairs.

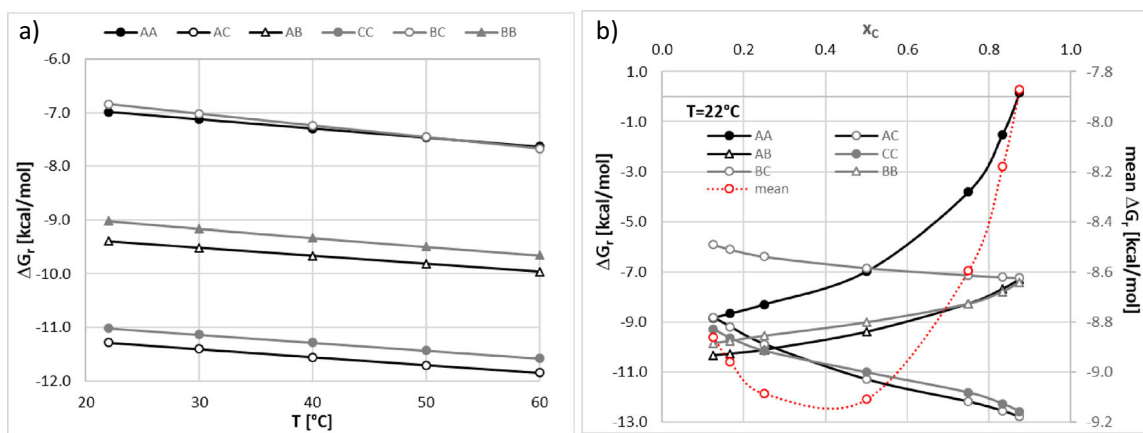

**Fig. 5** Quantification of the influence of temperature and amount of choline chloride (C) on the values of Gibbs free energy of binary intermolecular complexes formation. (**a**) NADES-curcumin 1:1 system (**b**) T = 22°C.

## CONCLUSIONS

Several important conclusions arise from the presented results. First of all, the solubility of curcumin in NADES was significantly larger than in the case of an aqueous solution. Among the studied formulations, the system with choline chloride and glycerol in unimolar proportions gave the highest solubility value, *i.e.* 7.25 mg/g in room temperature, which stands for an increase of 12 thousand times compared to aqueous solution, justifying the use of NADES as systems for efficient curcumin dissolution. This system was also tested as a potential medium for extraction of curcumin from natural sources and was quite successful in that role, particularly because of the fact that natural deep eutectic solvents are safe to use in food and pharmaceutical industries and can be administrated to individuals freely, which is usually not the case for organic solvents. This particular NADES was further tested as a stabilizing agent and it was found that it prevent the degradation of curcumin under the exposure to sunlight, contrary to methanol solution or even curcumin powder. Finally, the solubility of curcumin, both in the form of a powder and contained in NADES, was assessed in simulated gastrointestinal fluids, namely Fasted State Simulated Intestinal Fluid (FASSIF) and Fasted State Simulated Gastric Fluid (FASSGF). Based on these studies, it is evident that using NADES as an agent for curcumin delivery offers an advantage over administrating curcumin in the form of a powder, as the solubilities of curcumin pre-dissolved in NADES are 2 and 3.5 times higher in the two respective fluids, although for the optimal delivery of curcumin in NADES lower volumes of gastrointestinal fluids are preferable. Importantly, the solubility of curcumin in the intestinal fluid is significantly larger than in the gastric fluid, which suggests that this may be the preferred place for the delivery of curcumin as a drug.

The general conclusion drawn from the obtained results is that natural deep eutectic solvents can be utilized throughout the whole process of curcumin handling, *i.e.* from the extraction from natural sources up to oral administrations to individuals because of their advantageous properties as curcumin solvents, as well as environmental and health safety. It is also evident the topic of using NADES as agents for optimized delivery of active pharmaceutical ingredients certainly deserves a deeper insight, including clinical trials.

The performed quantum chemistry computations clearly indicated the origin of the enhanced solubility of curcumin in NADES in the presence of gastrointestinal fluids. Direct intermolecular contacts leading to hetero-molecular pairs with choline chloride and glycerol are responsible for elevating the bulk concentration of curcumin. Choline chloride plays a dominant role in the system and the complex formed with curcumin is the most stable among all possible homo- and hetero-molecular pairs that can be found in NADES-curcumin systems.

## ACKNOWLEDGMENTS AND DISCLOSURES

This work was supported by the grant number MN-1/WF/2017 for the development of young scientists in the Collegium Medicum, Nicolaus Copernicus University. Financial support is acknowledged.

**Open Access** This article is distributed under the terms of the Creative Commons Attribution 4.0 International License (http://creativecommons.org/licenses/by/4.0/), which permits unrestricted use, distribution, and reproduction in any medium, provided you give appropriate credit to the original author(s) and the source, provide a link to the Creative Commons license, and indicate if changes were made.

## REFERENCES

1. Basnet P, Skalko-Basnet N, Basnet P, Skalko-Basnet N. Curcumin: an anti-inflammatory molecule from a curry spice on the path to cancer treatment. Molecules. 2011;16(6):4567–9458.
2. Sugiyama Y, Kawakishi S, Osawa T. Involvement of the beta-diketone moiety in the antioxidative mechanism of tetrahydrocurcumin. Biochem Pharmacol. 1996;52(4):519–25.
3. Wang Y, Shi Y, Xu X, Liu F, Yao H, Zhai G, et al. Preparation of PANI-coated poly (styrene-co-styrene sulfonate) nanoparticles in microemulsion media. Colloids Surfaces A Physicochem Eng Asp. 2009;345(1–3):71–4.
4. Zhang T, Chen Y, Ge Y, Hu Y, Li M, Jin Y. Inhalation treatment of primary lung cancer using liposomal curcumin dry powder inhalers. Acta Pharm Sin B. 2018;8(3):440–8.
5. Gómez-Sierra T, Eugenio-Pérez D, Sánchez-Chinchillas A, Pedraza-Chaverri J. Role of food-derived antioxidants against cisplatin induced-nephrotoxicity. Food Chem Toxicol. 2018;120:230–42.
6. Parsamanesh N, Moossavi M, Bahrami A, Butler AE, Sahebkar A. Therapeutic potential of curcumin in diabetic complications. Pharmacol Res. 2018;136:181–93.
7. Shoba G, Joy D, Joseph T, Majeed M, Rajendran R, Srinivas P. Influence of Piperine on the pharmacokinetics of curcumin in animals and human volunteers. Planta Med. 1998;64(4):353–6.
8. Jamwal R. Bioavailable curcumin formulations: a review of pharmacokinetic studies in healthy volunteers. J Integr Med. 2018;16:367–74. https://doi.org/10.1016/J.JOIM.2018.07.001.
9. Mangolim CS, Moriwaki C, Nogueira AC, Sato F, Baesso ML, Neto AM, et al. Curcumin–β-cyclodextrin inclusion complex: stability, solubility, characterisation by FT-IR, FT-Raman, X-ray diffraction and photoacoustic spectroscopy, and food application. Food Chem. 2014;153:361–70.
10. Chuah AM, Jacob B, Jie Z, Ramesh S, Mandal S, Puthan JK, et al. Enhanced bioavailability and bioefficacy of an amorphous solid dispersion of curcumin. Food Chem. 2014;156:227–33.
11. Manju S, Sreenivasan K. Conjugation of curcumin onto hyaluronic acid enhances its aqueous solubility and stability. J Colloid Interface Sci. 2011;359(1):318–25.
12. Yang R, Zhang S, Kong D, Gao X, Zhao Y, Wang Z. Biodegradable polymer-curcumin conjugate micelles enhance the loading and delivery of low-potency curcumin. Pharm Res. 2012;29(12):3512–25.

13. O'Toole MG, Henderson RM, Soucy PA, Fasciotto BH, Hoblitzell PJ, Keynton RS, et al. Curcumin encapsulation in submicrometer spray-dried chitosan/tween 20 particles. Biomacromolecules. 2012;13(8):2309–14.
14. Anselmo AC, Mitragotri S. A review of clinical translation of inorganic nanoparticles. AAPS J. 2015;17(5):1041–54.
15. Williams HD, Trevaskis NL, Charman SA, Shanker RM, Charman WN, Pouton CW, et al. Strategies to address low drug solubility in discovery and development. Pharmacol Rev. 2013;65(1):315–499.
16. Takagi T, Ramachandran C, Bermejo M, Yamashita S, Yu L, Amidon G. A provisional biopharmaceutical classification of the top 200 Oral drug products in the United States, Great Britain, Spain, and Japan. Mol Pharm. 2006;3(6):631–43.
17. Kawabata Y, Wada K, Nakatani M, Yamada S, Onoue S. Formulation design for poorly water-soluble drugs based on biopharmaceutics classification system: basic approaches and practical applications. Int J Pharm. 2011;420(1):1–10.
18. Scholz A, Abrahamsson B, Diebold SM, Kostewicz E, Polentarutti BI, Ungell A-L, et al. Influence of hydrodynamics and particle size on the absorption of felodipine in labradors. Pharm Res. 2002;19(1):42–6.
19. Van Eerdenbrugh B, Van den Mooter G, Augustijns P. Top-down production of drug nanocrystals: Nanosuspension stabilization, miniaturization and transformation into solid products. Int J Pharm. 2008;364(1):64–75.
20. Przybyłek M, Cysewski P. Distinguishing Cocrystals from simple eutectic mixtures: phenolic acids as potential pharmaceutical Coformers. Cryst Growth Des. 2018;18(6):3524–34.
21. Korn C, Balbach S. Compound selection for development – is salt formation the ultimate answer? Experiences with an extended concept of the "100mg approach". Eur J Pharm Sci. 2014;57:257–63.
22. Blagden N, de Matas M, Gavan PT, York P. Crystal engineering of active pharmaceutical ingredients to improve solubility and dissolution rates. Adv Drug Deliv Rev. 2007;59(7):617–30.
23. Brewster ME, Loftsson T. Cyclodextrins as pharmaceutical solubilizers. Adv Drug Deliv Rev. 2007;59(7):645–66.
24. Huang L, Tong W-Q. Impact of solid state properties on developability assessment of drug candidates. Adv Drug Deliv Rev. 2004;56(3):321–34.
25. Janssens S, Van den Mooter G. Review: physical chemistry of solid dispersions. J Pharm Pharmacol. 2009;61(12):1571–86.
26. Kranz H, Guthmann C, Wagner T, Lipp R, Reinhard J. Development of a single unit extended release formulation for ZK 811 752, a weakly basic drug. Eur J Pharm Sci. 2005;26(1):47–53.
27. Feeney OM, Crum MF, McEvoy CL, Trevaskis NL, Williams HD, Pouton CW, et al. 50 years of oral lipid-based formulations: provenance, progress and future perspectives. Adv Drug Deliv Rev. 2016;101:167–94.
28. Abbott AP, Boothby D, Capper G, Davies DL, Rasheed RK. Deep eutectic solvents formed between choline chloride and carboxylic acids: versatile alternatives to ionic liquids. J Am Chem Soc. 2004;126(29):9142–7.
29. Choi YH, van Spronsen J, Dai Y, Verberne M, Hollmann F, Arends IWCE, et al. Are natural deep eutectic solvents the missing link in understanding cellular metabolism and physiology? Plant Physiol. 2011;156(4):1701–5.
30. Dai Y, Verpoorte R, Choi YH. Natural deep eutectic solvents providing enhanced stability of natural colorants from safflower (Carthamus tinctorius). Food Chem. 2014;159:116–21.
31. Duan L, Dou L-L, Guo L, Li P, Liu E-H. Comprehensive evaluation of deep eutectic solvents in extraction of bioactive natural products. ACS Sustain Chem Eng. 2016;4(4):2405–11.
32. Gutiérrez MC, Ferrer ML, Yuste L, Rojo F, del Monte F. Bacteria incorporation in deep-eutectic solvents through freeze-drying. Angew Chem Int Ed. 2010;49(12):2158–62.
33. Durand E, Lecomte J, Baréa B, Dubreucq E, Lortie R, Villeneuve P, et al. Evaluation of deep eutectic solvent–water binary mixtures for lipase-catalyzed lipophilization of phenolic acids. Green Chem. 2013;15(8):2275–82.
34. Li Z, Lee PI. Investigation on drug solubility enhancement using deep eutectic solvents and their derivatives. Int J Pharm. 2016;505(1–2):283–8.
35. Radošević K, Ćurko N, Gaurina Srček V, Cvjetko Bubalo M, Tomašević M, Kovačević Ganić K, et al. Natural deep eutectic solvents as beneficial extractants for enhancement of plant extracts bioactivity. LWT Food Sci Technol. 2016;73:45–51.
36. Paiva A, Craveiro R, Aroso I, Martins M, Reis RL, Duarte ARC. Natural deep eutectic solvents – solvents for the 21st century. ACS Sustain Chem Eng. 2014;2(5):1063–71.
37. Hayyan A, Mjalli FS, Al-Nashef IM, Al-Wahaibi YM, Al-Wahaibi T, Hashim MA. Glucose-based deep eutectic solvents: physical properties. J Mol Liq. 2013;178:137–41.
38. Silva LP, Fernandez L, Conceição JHF, Martins MAR, Sosa A, Ortega J, et al. Design and characterization of sugar-based deep eutectic solvents using conductor-like screening model for real solvents. ACS Sustain Chem Eng. 2018;6(8):10724–34.
39. COSMOlogic GmbH & Co. KG. COSMOconfX, version 4.1; 2018.
40. COSMOlogic GmbH & Co. KG. COSMOthermX, version 18.01; 2018.
41. COSMOlogic GmbH & Co. KG. Turbomole, version 7.0; 2015.
42. Hellweg A, Eckert F. Brick by brick computation of the gibbs free energy of reaction in solution using quantum chemistry and COSMO-RS. AICHE J. 2017;63(9):3944–54.
43. Tønnesen HH, Másson M, Loftsson T. Studies of curcumin and curcuminoids. XXVII. Cyclodextrin complexation: solubility, chemical and photochemical stability. Int J Pharm. 2002;244(1–2):127–35.
44. Schiller C, Frohlich C-P, Giesmann T, Siegmund W, Monnikes H, Hosten N, et al. Intestinal fluid volumes and transit of dosage forms as assessed by magnetic resonance imaging. Aliment Pharmacol Ther. 2005;22(10):971–9.